# Dial It Down: The Effect of Strongly Interacting Adsorbates on the $BiAg_2$ Rashba Surface State

Anubhab Chakraborty,[1] Torsten Fritz,[3] Percy Zahl,[4] Oliver L.A. Monti[1,2*]

[1]Department of Chemistry and Biochemistry, University of Arizona, Tucson, Arizona 85721, United States; email: anubhabc@arizona.edu

[2]Department of Physics, University of Arizona, Tucson, Arizona 85721, United States; email: monti@arizona.edu

[3]Friedrich Schiller University Jena, Institute of Solid State Physics, Helmholtzweg 5, 07743 Jena, Germany; email: torsten.fritz@uni-jena.de

[4]Brookhaven National Laboratory, Center for Functional Nanomaterials, Upton, New York 11973, United States; email: pzahl@bnl.gov

## Abstract

Organic semiconductors interfaced with spin-orbit coupled materials offer a rich playground for fundamental studies of controlling spin dynamics in spintronic devices. The adsorbate-surface interactions at such interfaces play a key role in determining the valence electronic and spin structure and consequently, the device physics as well. Here we present the adsorption and electronic structure of the strong organic electron acceptor 2,7-dinitropyrene-4,5,9,10-tetrone ($NO_2$-PyT, $C_{16}H_4N_2O_8$) on the Rashba spin-orbit coupled surface alloy $BiAg_2$/Ag(111). We show that the strong adsorbate-surface alloy interaction leads to weakening of the electronic coupling between the surface alloy atoms and quench the spin-orbit coupled surface state in $BiAg_2$/Ag(111). Our findings demonstrate an important challenge associated with using molecular adsorbates to tailor the spin polarization in $BiAg_2$/Ag(111), and our work provides guidelines to consider while designing interfacial systems to engineer the spin polarization in Rashba surface alloys.

## Introduction

The field of spintronics, which uses the electron's spin degree of freedom for information processing, has emerged as a strong candidate to be a building block for the next generation of electronics offering advantages like non-volatility, low power consumption, and faster operation.[1,2] A powerful way to achieve spin-based devices is making use of solid-state materials with intrinsic spin-orbit coupling (SOC) arising from structural inversion symmetry breaking, known as the Rashba SOC.[3–5] Recently it was discovered that surface alloying of noble metals with doped heavy metals leads to giant Rashba SOC in the surface electronic structure, some examples being Bi/Ag(111), Pb/Ag(111), Bi/Cu(111) and Sn/Ag(111).[6–11] The strong spin-splitting observed in the surface states of these systems have been attributed to both the potential gradient in the surface normal direction as well as the mixing of in-plane and out-of-plane orbitals due to inversion symmetry breaking.[12–14] These surface alloys therefore provide an excellent platform for manipulating spin texture and spin currents.

The integration of organic semiconductor layers interfaced with Rashba surface alloys opens up further exciting possibilities in this field. The Rashba effect at the metal-organic interface can in principle induce spin-momentum locking in the organic layer, and combined with the long spin lifetimes of organic semiconductors[15] can potentially facilitate spin injection and transport to create organic spintronic devices.[16–21] Recent works have explored interface formation and the effect of molecular adsorption on the structural and electronic properties of Rashba surface alloys.[10,22–26] Based on these works, it has been proposed that formation of localized $\sigma$-bonds between the adsorbate layer and the atoms of the surface alloy can enhance the Rashba spin-splitting in these materials,[10,26] which can allow control over spin polarization without external magnetic fields. However, this empirical rule is yet to be validated experimentally.

In this work, we study the adsorption of the strong organic electron acceptor 2,7-dinitropyrene-4,5,9,10-tetrone ($NO_2$-PyT, $C_{16}H_4N_2O_8$) on the Rashba surface alloy $BiAg_2$/Ag(111). The presence of -$NO_2$ and -C=O functional groups in this molecule provides multiple sites for local $\sigma$-bond formation with the surface alloy atoms and therefore is an ideal system to study potential Rashba SOC enhancement using adsorbate interaction. Using Ultraviolet Photoemission Spectroscopy (UPS) and Angle-Resolved Photoemission Spectroscopy (ARPES), we demonstrate the challenges associated with this approach of tailoring the Rashba SOC in surface alloys. We show that strongly interacting adsorbate-surface alloy systems can in fact lead to weakening of the electronic coupling between the surface alloy atoms and quench the Rashba surface state in these materials. Our results answer the question whether it is sufficient to interface a Rashba surface alloy with a strongly interacting adsorbate having $\sigma$-bonding capabilities, and our conclusions therefore provide additional guidelines for designing interfacial structures of Rashba SOC materials, with implications for organic spintronic device engineering.

## Methods

$NO_2$-PyT was commercially obtained (Alfa Chemical, 97%) and purified by three cycles of gradient sublimation (500 K) in a custom-built vacuum furnace ($5 \times 10^{-6}$ Torr). The Ag(111) crystal was cleaned using repeated cycles of $Ar^+$ sputtering (1 keV, 5 − 10 μA $cm^{-2}$) and annealing (800 K). Bi was deposited onto Ag(111) in a UHV sample preparation chamber ($7 \times 10^{-10}$ Torr) using a custom-built water cooled Knudsen source, and the deposition rate (0.05 ML $min^{-1}$) was monitored using a quartz crystal microbalance (QCM). Bi coverage on Ag(111) was determined using work function and LEED pattern analysis.[27] $NO_2$-PyT was deposited using a custom-built Knudsen source at a rate of (0.01-0.04 ML $min^{-1}$). $NO_2$-PyT coverage on Bi/Ag(111) was estimated using work function[28] and Low-Energy Electron Diffraction (LEED) analysis. All UPS, ARPES and LEED measurements were performed under UHV conditions ($2 \times 10^{-10}$ Torr) and at 298 K. The UPS and ARPES measurements were taken in a VG EscaLab MK II photoelectron spectrometer using a He I photon source (SPECS 10/35, $h\nu$ = 21.22 eV), with an analyzer acceptance angle of $\pm 1.5°$ and a sample bias of -5 V. LEED images were acquired using an Omicron SPECTALEED instrument, and image analysis and distortion correction were done using LEEDCal and LEEDLab.[29] High Resolution Atomic Force Imaging (HR-AFM) measurements were performed with a Createc-based low-temperature (LT)STM system custom upgraded with HR-AFM capability and operated using the open source GXSM control software.[30,31] A small bias of 20 mV was typically applied in HR-AFM-mode. For frequency detection, the custom, high-speed GXSM RedPitaya-PAC-PLL controller was used.

## Results

### *Growth of Bi on Ag(111)*

The Bi/Ag(111) surface structure is well-studied,[6,9,27,32,33] and we start by discussing the surface structure evolution as a function of Bi coverage using our Low-Energy Electron Diffraction (LEED) results (Figure 1) that are in strong agreement with literature. The growth of Bi on Ag(111) can be categorized into three phases: (1) At low coverages ($\theta \leq 0.55$ ML), a $BiAg_2$ surface alloy phase forms due to incorporation of Bi atoms on the topmost Ag layer by substitution, as well as due to chemical bonding of Bi atoms with Ag atoms released during substitution to form $BiAg_2$ surface alloy islands on the Ag(111) surface.[32] In this phase, the Bi atoms protrude out of the Ag plane by $d = 0.65$ Å, known as the corrugation parameter.[34,35] The surface alloy forms a $(\sqrt{3} \times \sqrt{3})R30°$ superstructure on Ag(111) ($\mathbf{a_{Ag}} = 2.89$ Å), as shown in Figure 1(a). (2) At higher coverages up to 1 ML (0.55 ML < $\theta \leq 1$ ML), a de-alloying process occurs in order to relieve the compressive strain induced by further incorporation of Bi atoms on the Ag surface.[32,36] The result of this de-alloying process is to convert the surface alloy phase to a Bi superstructure with a $(p \times \sqrt{3})$ periodicity, where this superstructure is only commensurate in the Ag$[11\bar{2}]$ direction. Our LEED results (Figure 1(b)-(c)) show the 2D reciprocal unit cell corresponding to this phase, and based on the LEED image analysis we estimate the following lattice parameters for the Bi- $(p \times \sqrt{3})$ rectangular unit cell: $\mathbf{a} = 4.42(2)$Å, $\mathbf{b} = 4.99(2)$Å $\approx \sqrt{3}\mathbf{a_{Ag}}$, $\gamma = 90°$, which are in reasonable agreement with literature.[32](3) For even higher coverages ($\theta > 1$ ML), the de-alloying transition is complete and Bi(110) thin films form on the surface in a rectangular unit cell, with decreased lattice constants compared to the $(p \times \sqrt{3})$ unit cell.[32] Our LEED results (Figure 1(d)) show the expected rectangular 2D reciprocal lattice, although we did not observe any significant change in the lattice constants. Even though the spin-split Rashba Surface State is only associated with the $(\sqrt{3} \times \sqrt{3})R30°$ $BiAg_2$ surface alloy phase, it is important to investigate the surface structure of higher coverage Bi/Ag(111) to understand the adsorbate-induced changes to the surface electronic structure, as explained later in this work.

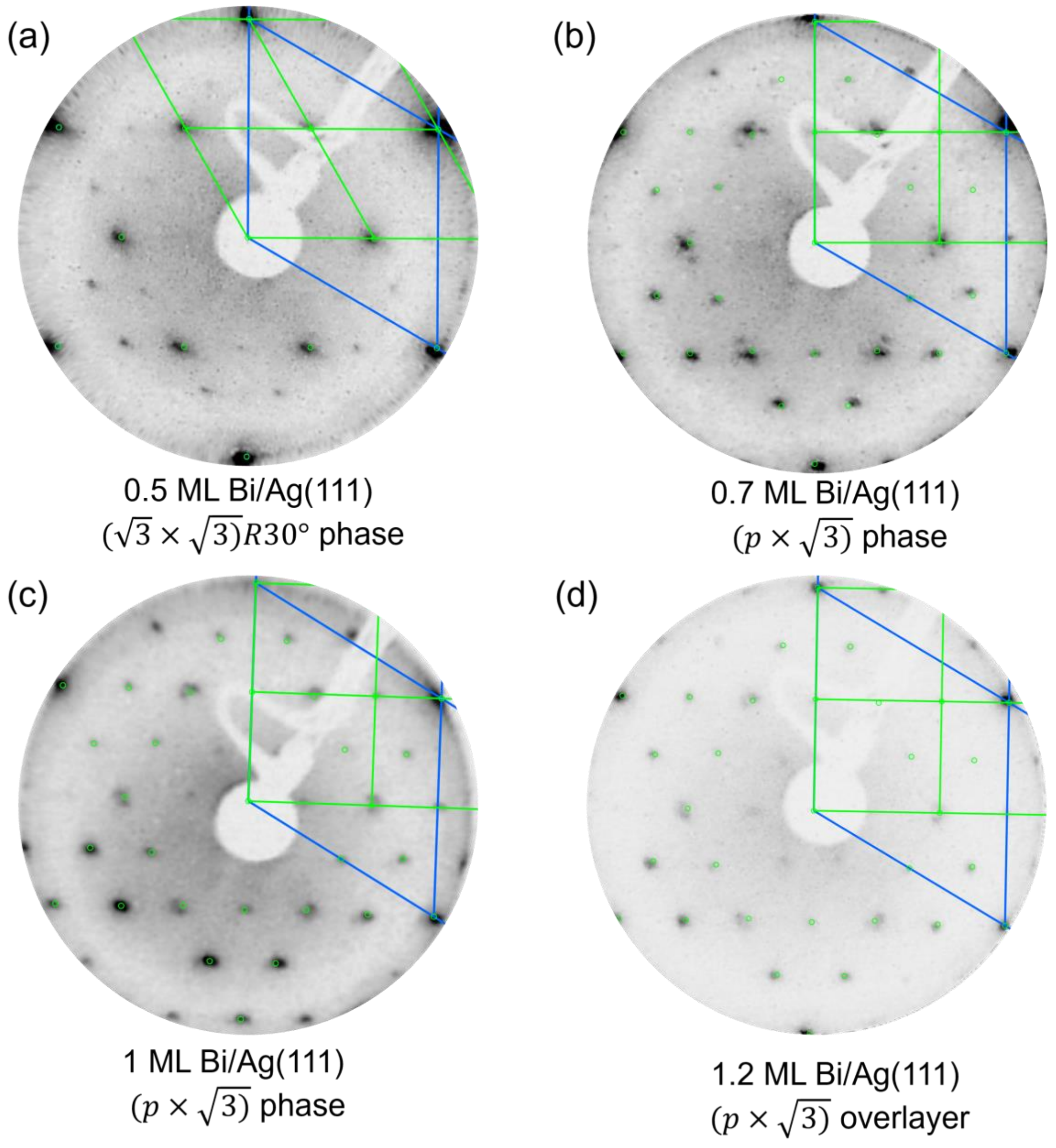

**Figure 1. Contrast-inverted LEED image of (a) 0.5 ML Bi/Ag(111), (b) 0.7 ML Bi/Ag(111), (c) 1 ML Bi/Ag(111) and (d) 1.2 ML Bi/Ag(111) acquired at 42.5 eV ((a), (c)-(d)) and 43 eV((b)). Lines indicate the 2D reciprocal unit cell of Ag(111) (blue) and that of the Bi superstructure (green), showing the evolution of the surface structure with increasing Bi coverage.**

*Growth of $NO_2$-PyT on Bi/Ag(111)*

Next, we discuss the adsorption structure of $NO_2$-PyT on the Bi/Ag(111) surfaces. This pyrene-derived acceptor (Figure 2(a)) has been studied on the Ag(111) surface,[28] where the following adsorption mechanism was proposed: the $NO_2$-PyT molecules initially adsorb in a “face-on” (also called “flat-lying”) orientation to form a monolayer on Ag(111). Upon further deposition, a rearrangement from a “face-on” monolayer to “edge-on” monolayer occurs at room temperature until surface saturation is reached with this orientation. Such a density-dependent reorientation has also been observed in the organic acceptor HATCN on Ag(111), attributed to the higher molecule-substrate interaction achieved in this edge-on orientation at

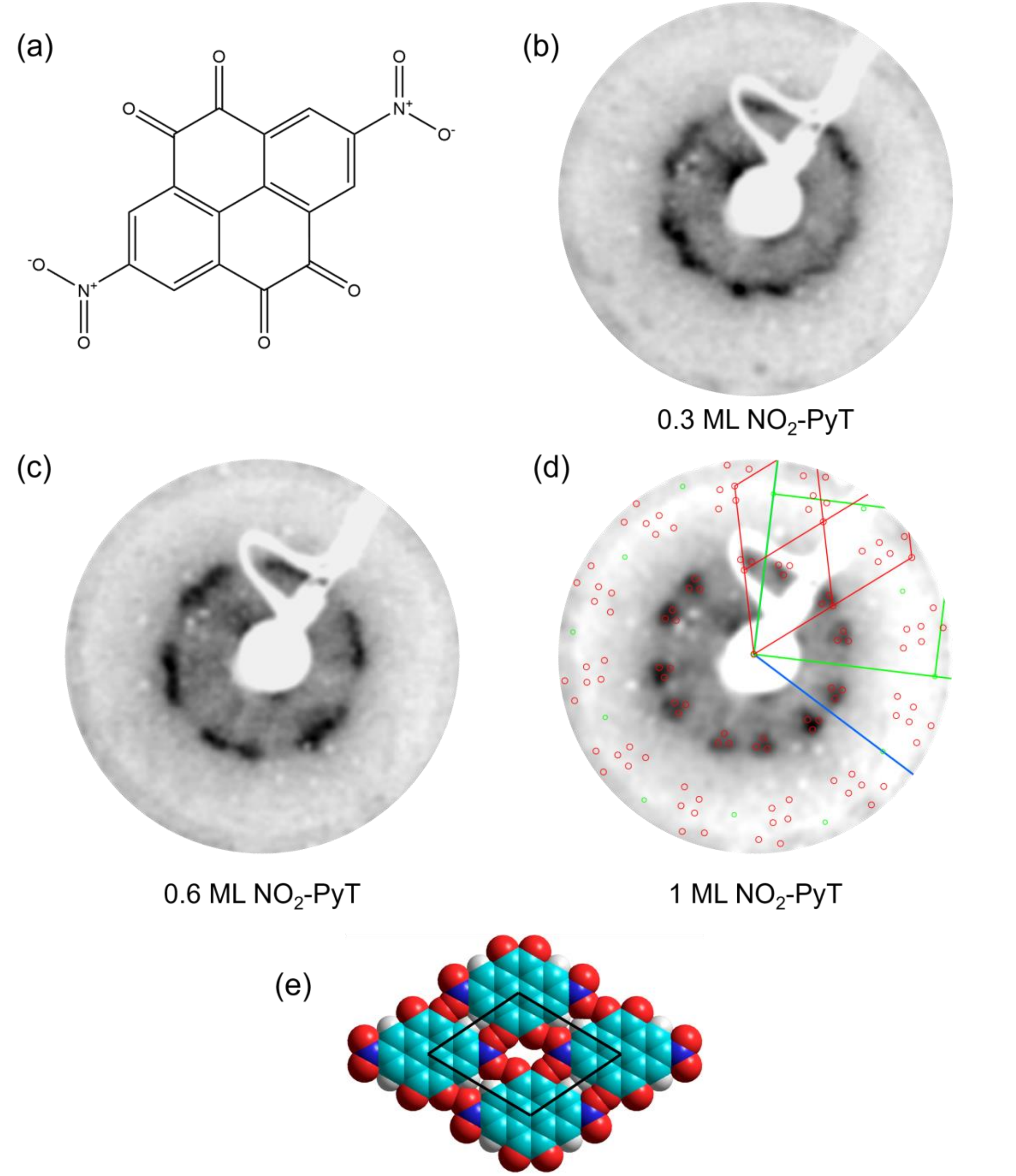

**Figure 2. (a) Molecular structure of $NO_2$-PyT. Contrast-inverted LEED images acquired at 15 eV of (b) 0.3 ML $NO_2$-PyT, (c) 0.6 ML $NO_2$-PyT and (d) 1 ML $NO_2$-PyT on 1.2 ML Bi/Ag(111). Lines indicate the 2D reciprocal unit cell of Ag(111) (blue), the Bi- (p×√(3)) rectangular unit cell (green) and that of the $NO_2$-PyT superstructure (red). (e) Proposed unit cell structure of $NO_2$-PyT/1.2 ML Bi/Ag(111) based on the LEED pattern analysis. Some higher-order diffraction spots are too weak to be observed.**

sufficient molecule densities on the surface.[37,38] Our High-Resolution AFM (HR-

AFM) imaging of sub-monolayer $NO_2$-PyT/Ag(111) (Figure S1, Supplementary Information) confirms the "face-on" adsorption orientation at low coverages.

We investigated the surface structure of $NO_2$-PyT on Bi/Ag(111) surfaces using LEED. No clear LEED pattern was observed for 1 ML $NO_2$-PyT on the $BiAg_2$ $(\sqrt{3} \times \sqrt{3})R30°$ surface alloy (i.e., 0.5 ML Bi/Ag(111)) to reveal any useful information about the adsorption structure. It is possible that the size of $BiAg_2$ surface alloy islands (~50 nm) [32,39] is not large enough to facilitate long-range ordering of the $NO_2$-PyT molecules on this surface. On the other hand, LEED of $NO_2$-PyT deposited on the Bi- $(p \times \sqrt{3})$ overlayer (i.e., 1.2 ML Bi/Ag(111)) reveals important features regarding the adsorption structure, as shown in Figure 2. We observe the formation of a LEED pattern which becomes more distinct with increasing $NO_2$-PyT coverage (Figure 2(b)-(d)) until the formation of 1 ML of $NO_2$-PyT. Due to the small number of spots, we could not fit the LEED data and instead manually determined the lattice parameters that explain the LEED pattern the best. Within these limitations, we find that the following lattice parameters explain the LEED pattern (Figure 2(d)) reasonably well: $\mathbf{a} = 9.6\ \text{Å},\ \mathbf{b} = 10.5\ \text{Å},\ \gamma = 115°, \varphi = 13.5°$, where φ is the angle to the Bi- $(p \times \sqrt{3})$ structure. No clearly discernible epitaxial registry of the $NO_2$-PyT structure was found to either the Bi- $(p \times \sqrt{3})$ or the Ag(111) structure from LEED analysis. Based on these lattice parameters, we propose a tentative oblique unit cell structure for $NO_2$-PyT/1.2 ML Bi/Ag(111), as illustrated in Figure 2(e).

The LEED pattern, which consists of six triple spots (red circles, Figure 2(d)), can be explained by considering a superposition of six symmetry-equivalent domains of the $NO_2$-PyT unit cell on the surface. The number of molecular domains is a direct consequence of the number of Bi domains on Ag(111). As the Bi- $(p \times \sqrt{3})$ overlayer forms a rectangular lattice with one Bi lattice vector (**b**) aligned with a primitive Ag(111) lattice vector, there are 3 rotationally equivalent domains of Bi (see Figure 1(b)-(d)) due to the three-fold symmetry of Ag(111) in real space. The $NO_2$-PyT molecules can grow on each of those domains (either Bi or Ag), hence forming 3 rotational domains. Furthermore, as the symmetry of the $NO_2$-PyT lattice is different from both Bi and Ag(111) and neither of the two primitive lattice vectors are aligned with either Bi- $(p \times \sqrt{3})$ or Ag(111) lattice vectors, mirror domains are present as well. These underlying symmetry and structural factors therefore give rise to the observed six triple spots. Interestingly, for >1 ML coverage of $NO_2$-PyT, the distinct LEED pattern continuously weakens and disappears completely for ~2 ML $NO_2$-PyT coverage. A likely explanation for this observation is the density-dependent reorientation of $NO_2$-PyT on Bi/Ag(111) to an "edge-on" orientation that is expected to alter the adsorption structure associated with the "face-on" orientation. Such a disappearance of the LEED pattern was observed for > 1ML HATCN/Ag(111), also likely due to density-dependent reorientation of the HATCN molecules.[38] An STM/AFM analysis of the surface, not part of this study, is needed to ascertain the validity of this explanation.

*Electronic valence band structure*

The spin-orbit coupled Rashba Surface State (RSS) of the $(\sqrt{3} \times \sqrt{3})R30°$ $BiAg_2$ surface alloy exists in the $L$-projected bulk band gap of Ag(111) and exhibits a large Rashba splitting parameter ($\alpha_R \approx 3.2$ eV Å) which is significantly higher than both Ag(111) and Bi(111).[6,8,40] The large spin splitting is attributed to the localization of the surface state in the surface layer as well as due to the outward relaxation of the Bi atoms, creating a corrugated surface with broken inversion symmetry.[8,34] The energy dispersion of the RSS is given by:

$$E(\boldsymbol{k}_{\parallel}) = \frac{\hbar^2}{2m^*}(k_{\parallel} \pm k_0)^2 + E_0 \tag{1}$$

The experimental RSS band dispersion is shown in Figure 3(a) and consists of a pair of parabolic bands symmetrically shifted about the $\Gamma$ point by a momentum offset $k_0 = 0.13$ Å$^{-1}$, with a binding energy $E_0 = -0.18$ eV and an effective mass $m^* = -0.3m_e$ obtained from fits using (1), which are in good agreement with literature.[6] This fully occupied RSS has mostly Bi $(s, p_z)$ character, with some Ag $s$ contribution as well.[7,8,26,34]As the Bi coverage on Ag(111) is increased, an evolution of the band structure is observed (Figure 3(b)-3(d)). For 0.7 ML Bi/Ag(111), both the $(\sqrt{3} \times \sqrt{3})R30°$ surface alloy and the Bi- $(p \times \sqrt{3})$ phase co-exist on the surface[9,32] and hence the RSS is still observed in ARPES (Figure 3(b)). The RSS disappears upon deposition 1 ML Bi/Ag(111) (Figure 3(c)) as the de-alloying process is complete and we observe new features in the band structure at this coverage and higher (1.2 ML Bi/Ag(111)) (Figure 3(d)), which are most likely Bi $6p$ states.[41–43]

Next, we discuss the effect of $NO_2$-PyT adsorption on the valence electronic structure of the $(\sqrt{3} \times \sqrt{3})R30°$ $BiAg_2$ surface alloy. The work function of the clean

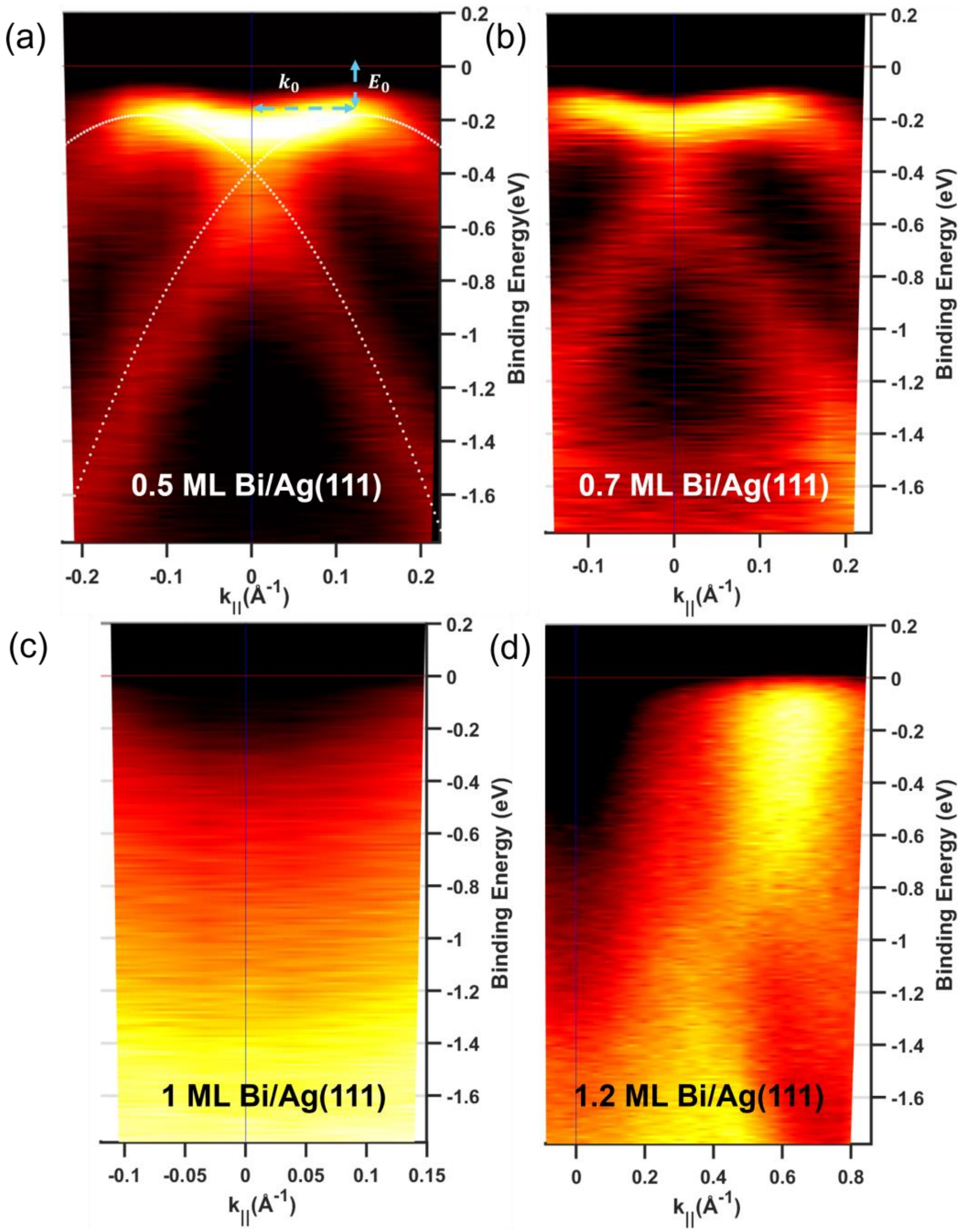

**Figure 3. ARPES in the *Γ→K* direction (of $BiAg_2$ (√3×√(3)) *R*30° superstructure) of (a) 0.5 ML Bi/Ag(111), (b) 0.7 ML Bi/Ag(111), (c) 1 ML Bi/Ag(111) and (d) 1.2 ML Bi/Ag(111). The dotted white curves in (a) represent the parabolic fit to the experimental data using equation (1).**

Ag(111) and the $BiAg_2$ surface alloy (0.5 ML Bi/Ag(111)) surfaces are 4.5 eV and 4.37 eV respectively. Upon deposition of $NO_2$-PyT, a large increase in the work function is observed (Figure S2) and reaches 5.32 eV at an estimated 1 ML deposition. The maximum work function of 5.5 eV is reached at ~ 2 ML and further deposition leads to a gradual decrease in the work function. The work function evolution is similar to $NO_2$-PyT/Ag(111), and indicates a charge transfer from the surface to the molecule.[28]The increase in the work function from 1 ML to 2 ML coverage is likely due to the re-orientation of the $NO_2$-PyT molecules to an edge-on orientation.

Most importantly, our UPS (Figure 4(a)-(b)) and ARPES (Figure 4(c)-(f)) results show significant changes in the electronic structure due to $NO_2$-PyT adsorption. Firstly, the $BiAg_2$ RSS (shown in Figure 3(a)) intensity decreases continuously with increasing $NO_2$-PyT coverage (Figure 4(c)-(d)) and disappears completely upon ~0.6 ML $NO_2$-PyT deposition (Figure 4(e)). In this process, no significant modifications to the RSS parameters (equation (1)) such as the $E_0$ were observed. Second, multiple new spectral features are observed. Comparing these features with the spectral features observed in $NO_2$-PyT/Ag(111),[28] we find several similarities and this allows us to assign the following: (1) The feature at -1 eV (Figure 4(b), (e)) is an Interfacial Charge-Transfer state (ICT) which shows a flat dispersion and is associated with an electronic charge transfer into the $NO_2$-PyT LUMO, (2) the broad feature at ~-3 eV (black arrow, Figure 4(a), Figure 4(f)) is associated with the former HOMO of $NO_2$-PyT and (3) the features at ~-9 eV and ~-10.5 eV are associated with deeper-lying $NO_2$-PyT states bound to the substrate. Of particular interest is the ICT state at -1 eV that is apparently due to a charge transfer from the 0.5 ML Bi/Ag(111) substrate into the $NO_2$-PyT LUMO. Similar charge-transfer induced ICT states have also been reported for other organic electron acceptor adsorbates on metal surfaces in previous works.[44–50] As seen on Figure 4(b), the ICT state achieves maximum intensity at 1 ML $NO_2$-PyT and decreases in intensity with further coverage which highlights the interfacial nature of this state. Interestingly, the spectral shape and binding energy of the $NO_2$-PyT/0.5 ML Bi/Ag(111) ICT state are significantly different than the $NO_2$-PyT/Ag(111) ICT state (Figure 5(a)). To understand this difference as well as the reason for the disappearance of the RSS, we further investigate the orbital composition of this ICT state.

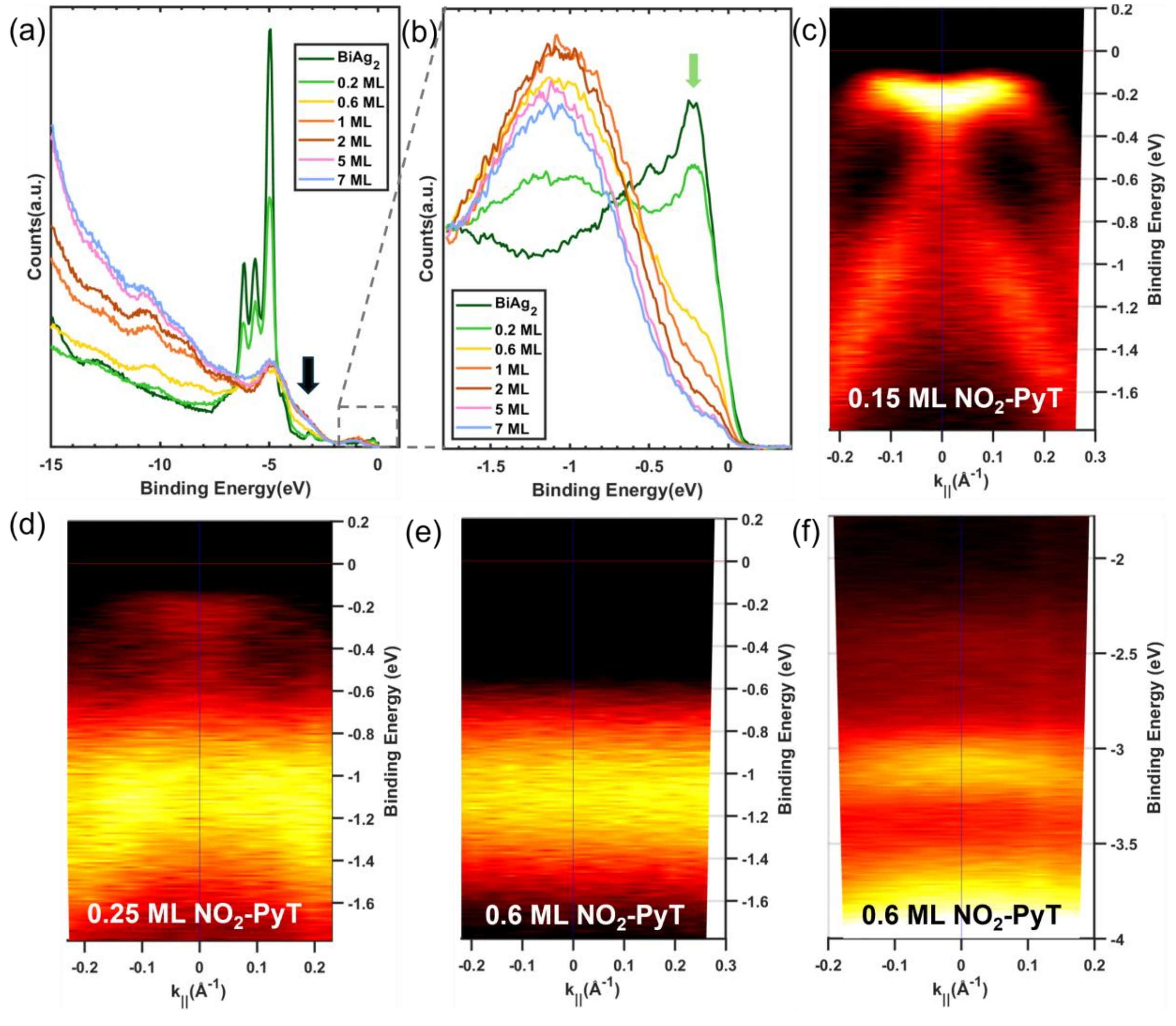

**Figure 4. UPS at normal emission ($k_{\|} = 0$) of $NO_2$-PyT/0.5 ML Bi/Ag(111) (a)-(b), showing (a) the full valence spectra and (b) zoom-in to the region near $E_F$ (grey dashed rectangle in (a)) showing the Interfacial Charge-Transfer (ICT) state. The black arrow in (a) shows the $NO_2$-PyT HOMO and the green arrow in (b) shows the RSS. The $NO_2$-PyT coverages (in ML) are indicated in the legend. ARPES in the $\Gamma \rightarrow K$ direction (of $BiAg_2$ (√3×√(3)) $R30°$ superstructure) of $NO_2$-PyT/0.5 ML Bi/Ag(111) for (c) 0.15 ML $NO_2$-PyT (d) 0.25 ML $NO_2$-PyT and (e), (f) 0.6 ML $NO_2$-PyT. (e) and (f) show the region near the ICT state and the $NO_2$-PyT HOMO respectively.**

*Origin of the ICT state*

To understand the nature of the ICT state on the $BiAg_2$ surface alloy substrate (i.e., 0.5 ML Bi/Ag(111)), it is useful to examine the nature of interaction of the $NO_2$-PyT molecules with a Ag surface as well as with a Bi surface. Therefore, we studied the valence electronic structures of $NO_2$-PyT/Ag(111) and $NO_2$-PyT/1.2 ML Bi/Ag(111) (Figure 5). The valence band spectrum for Ag(111) (dark green curve, Figure 5(a)) contains the characteristic Shockley Surface State near $E_F$.[51,52]Upon $NO_2$-PyT deposition, a charge transfer of ~0.7 electrons[28] occurs into the $NO_2$-PyT LUMO due to hybridization with Ag states near $E_F$ to form the ICT state. This state appears in the UPS as an additional intensity right below $E_F$. The spectral shape can be understood as a asymmetric Gaussian feature arising from a partially occupied former-LUMO, cut off by the $E_F$.[48,53] This indicates a metallic character of the $NO_2$-PyT/Ag(111) ICT state. On the other hand, the ICT state of $NO_2$-PyT/ $BiAg_2$ lies at a higher binding energy, well below $E_F$ (Figure 4(b)). This is indicative of a greater charge transfer to the $NO_2$-PyT LUMO on the $BiAg_2$ surface, which creates a fully-occupied and semiconducting ICT state.[54] These observations suggest a significant difference in the hybridization mechanism of the $NO_2$-PyT LUMO on a Ag(111) surface compared to a $BiAg_2$ surface. The scenario is markedly different for $NO_2$-PyT/1.2 ML Bi/Ag(111), which is a fully Bi-covered surface with a Bi(110) surface structure,[32] as described earlier in this work. On this surface, the density of states (DOS) near $E_F$ are due to the Bi $6p$ states (dark green curve, Figure 5(b)-(c) and Figure 3(d)), including dangling $6p_z$ states.[55]The results upon $NO_2$-PyT deposition on this surface are very similar to $NO_2$-PyT/$BiAg_2$. The broad feature associated with the former-HOMO ~-3 eV is observed (Figure 5(c)) in UPS, as well as the formation of an ICT state at lower binding energies (Figure 5(b)). The ICT state binding energy is -1.10(1) eV for 1 ML $NO_2$-PyT and it increases to -1.34(1) eV for 2 ML $NO_2$-PyT coverage (Figure 5(b), dashed lines). This binding energy shift is most likely due to the reorientation of the molecules to an edge-on orientation between 1 and 2 ML coverage, and a similar effect has been observed on other organic adsorbate-on-metal systems as well.[28,56,57]Importantly, the spectral shape of the ICT states of $NO_2$-PyT/$BiAg_2$ and $NO_2$-PyT/1.2 ML Bi/Ag(111) are remarkably similar (yellow and orange curves, Figure 5(d)) and the binding energy difference of 100 meV is possibly due to the work function (and vacuum level) shift between the $BiAg_2$ and the 1.2 ML Bi surfaces by the same amount. This strongly indicates that the $NO_2$-PyT LUMO primarily interacts with the Bi $6p$ states near $E_F$ and consequent charge-transfer leads to the formation of the $NO_2$-PyT/$BiAg_2$ ICT state. Having established the origin of the ICT state, we will discuss the implications of our findings next.

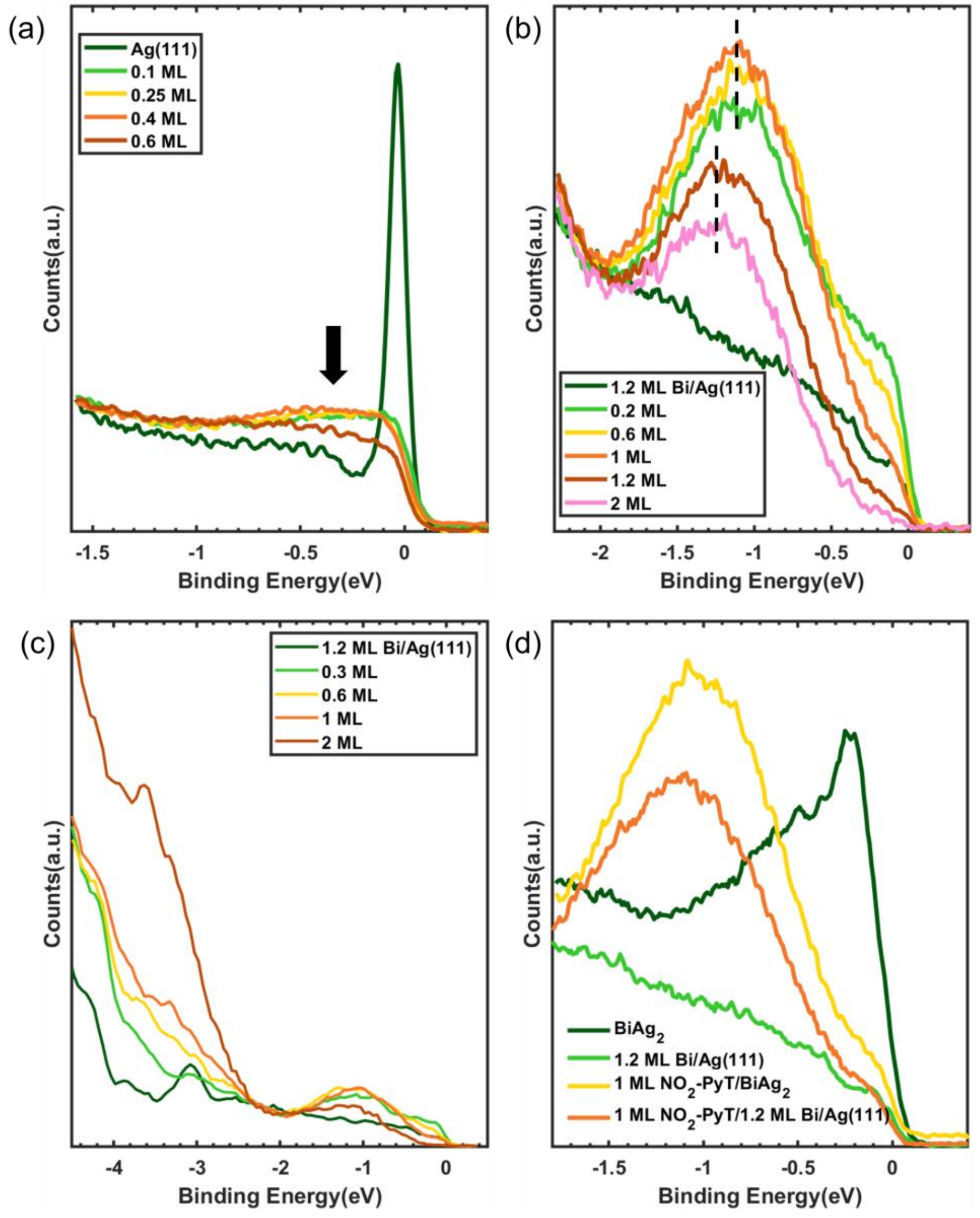

**Figure 5. UPS at normal emission ($k_{\parallel} = 0$) of (a) $NO_2$-PyT/Ag(111) and (b)-(c) $NO_2$-PyT/1.2 ML Bi/Ag(111), showing the regions containing the ICT state and the former-HOMO, respectively. The black arrow in (a) shows the additional intensity associated with the ICT state. The dashed lines in (b) show the binding energy shift of the ICT state between 1 ML $NO_2$-PyT and 2 ML $NO_2$-PyT coverage. The $NO_2$-PyT coverages (in ML) are indicated in the legend. (d) Comparison of the ICT state of $NO_2$-PyT/$BiAg_2$ (yellow) with the ICT state of $NO_2$-PyT/1.2 ML Bi/Ag(111) (orange).**

*Discussion*

Our results bear strong resemblance to two previous works on organic acceptors on metal surfaces: NTCDA/0.1 ML Na/Ag(111)[54] and PTCDA/$PbAg_2$ surface alloy[23]. In the former case, a similar transition from a metallic ICT state on NTCDA/Ag(111) to a semiconducting ICT state on NTCDA/0.1 ML Na/Ag(111) was observed. This was attributed to the strong coupling and charge transfer between NTCDA and on-surface Na atoms which in turn decoupled the NTCDA molecules and the Na atoms to the underlying Ag(111) surface, thereby changing the nature of the ICT state. In the latter case, PTCDA forms $\sigma$-bonds with the Pb atoms of $PbAg_2$, leading to a vertical displacement of the Pb atoms which causes de-alloying and vanishing of the $PbAg_2$ RSS. The formation of an ICT state was also observed due to this hybridization. Therefore, the physical picture that arises for $NO_2$-PyT adsorption on $BiAg_2$ in analogy is as follows: upon adsorption, the $NO_2$-PyT LUMO interacts strongly and selectively with the surface Bi $6p$ states near $E_F$ and leads to a charge-transfer to form a semiconducting ICT state. This hybridization and charge-redistribution weakens the Bi-Ag coupling, leading to the disappearance of the RSS in the valence band spectra.

We contextualize our findings within the broader scope of modifying Rashba spin-orbit coupling in surface alloys using adsorbates. The $BiAg_2$ RSS is known to be quite robust to any modification by organic overlayers,[10,22] which was attributed to the fully-occupied nature of the $BiAg_2$ RSS as well as the absence of $\sigma$-bond formation in such adsorbed systems. In an earlier computational work,[26] it was proposed that adsorbates can be used to enhance the outward buckling of the Bi atoms ($d > 0.65$ Å) that increases the Rashba splitting magnitude in $BiAg_2$. In this regard, using organic acceptors with functional groups that can form localized $\sigma$-bonds with the surface and increase the outward buckling of Bi is considered to be a viable way of modifying the Rashba splitting in $BiAg_2$. Our work highlights the limitations of this approach. We show that a sufficiently strong adsorbate-Bi coupling can in fact lead to electronic decoupling of Bi and Ag atoms in the surface alloy and instead of an enhancement, it can quench the RSS altogether. In the vein of a previous work,[10] we therefore propose an empirical rule to consider while designing an interfacial system to engineer the RSS: for strongly interacting adsorbates, it is important to control the extent of the adsorbate-surface alloy electronic coupling to prevent quenching of the RSS. This can in principle be achieved by functional group engineering of molecular adsorbates.[23,26]

*Conclusions*

In this work, we investigated the structure and valence electronic properties of $NO_2$-PyT adsorbed on Bi/Ag(111) surfaces. We show that $NO_2$-PyT forms an ordered structure on Bi/Ag(111) and likely undergoes a density-dependent reorientation to an "edge-on" orientation for > 1 ML coverage. We observe the formation of an ICT state due to $NO_2$-PyT and Bi interaction and charge-transfer. This significantly affects the surface electronic structure and leads to quenching of the RSS in the $BiAg_2$ surface alloy. Based on our findings, we propose that for strongly interacting adsorbates, it is important to control the extent of the adsorbate-surface alloy electronic coupling to prevent quenching of the RSS. Our work highlights the effect of strong interactions on the surface electronic structure and provides a strategy for better interfacial system design to engineer the Rashba splitting in $BiAg_2$ and related surface alloys.

For Table of Contents Only

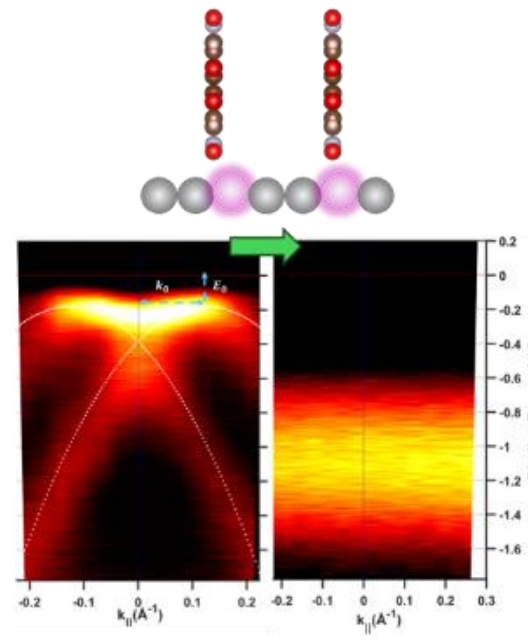

Supporting Information: Dial It Down: The Effect of Strongly Interacting Adsorbates on the $BiAg_2$ Rashba Surface State

Anubhab Chakraborty,[1] Torsten Fritz,[3] Percy Zahl,[4] Oliver L.A. Monti[1,2*]

[1]Department of Chemistry and Biochemistry, University of Arizona, Tucson, Arizona 85721, United States; email: anubhabc@arizona.edu

[2]Department of Physics, University of Arizona, Tucson, Arizona 85721, United States; email: monti@arizona.edu

[3]Friedrich Schiller University Jena, Institute of Solid State Physics, Helmholtzweg 5, 07743 Jena, Germany; email: torsten.fritz@uni-jena.de

[4]Brookhaven National Laboratory, Center for Functional Nanomaterials, Upton, New York 11973, United States; email: pzahl@bnl.gov

S1: HR-AFM results of 0.1 ML $NO_2$-PyT/Ag(111)

At low coverages ($\theta$< 1 ML), the $NO_2$-PyT molecules adsorb on the Ag(111) surface in a flat-lying (or face-on) orientation. This was confirmed by our HR-AFM images for 0.1 ML $NO_2$-PyT/Ag(111) (Figure S1).

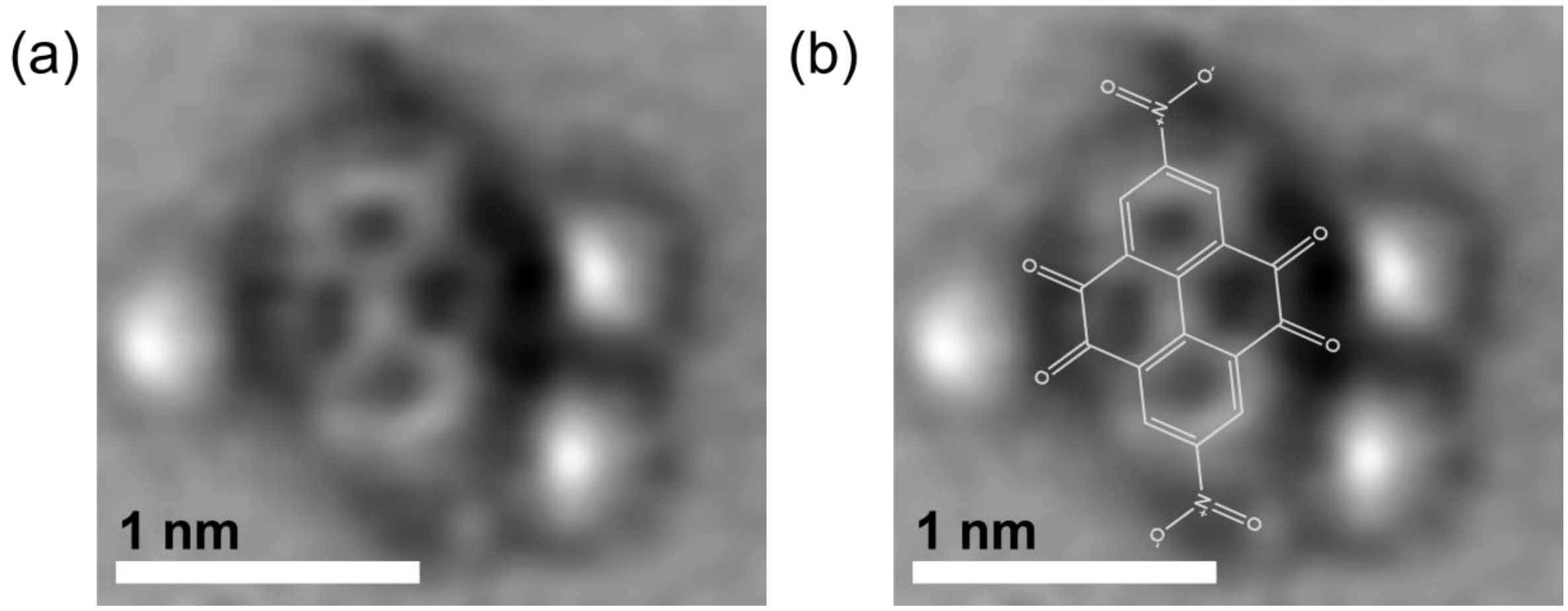

**Figure S1. Constant height HR-AFM (0.02 V) of 0.1 ML NO2-PyT/Ag(111). The three bright spots surrounding the molecule are most likely CO molecules adsorbed on Ag(111). (b) Same as (a), but the molecular structure overlaid.**

S2: Work Function trend of $NO_2$-PyT/0.5 ML Bi/Ag(111)

The work function of the Rashba surface alloy (0.5 ML Bi/Ag(111)) is 4.37 eV, and decreases to 4.2 eV for 1.2 ML Bi/Ag(111), as expected from pervious works.[1] When $NO_2$-PyT is deposited on the Rashba surface alloy, the work function increases in a similar trend as reported for $NO_2$-PyT/Ag(111),[2] reaching a maximum value of 5.5 eV for 2 ML coverage and shows gradual decrease for even higher coverages. We combine the work function and LEED results (Figure 2, main text) to estimate the $NO_2$-PyT coverage on the surface.

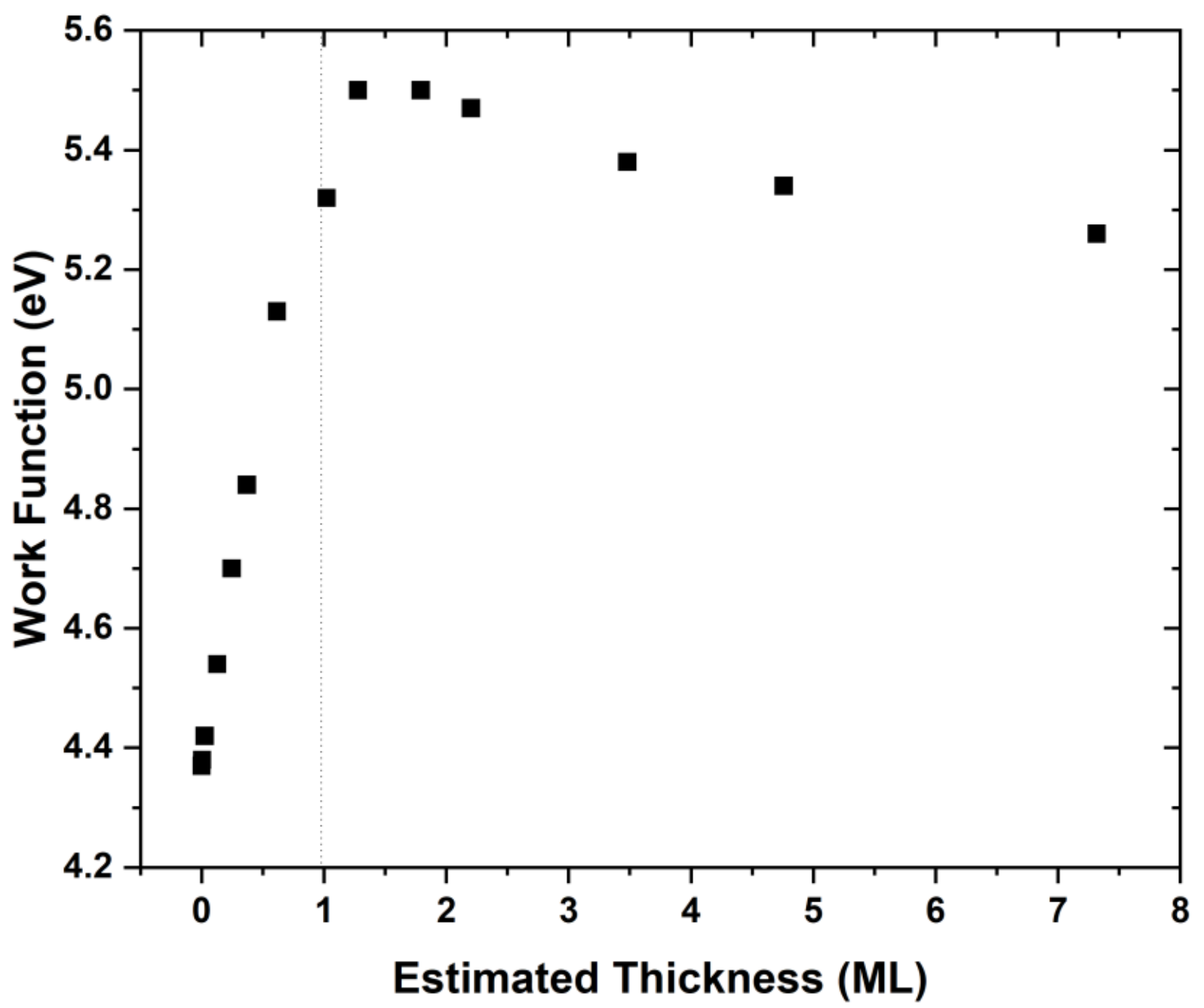

**Figure S2. Work Function as a function of NO2-PyT coverage (in ML) on the 0.5 ML Bi/Ag(111) surface.**